\documentclass[pra,aps, showpacs, twocolumn, a4paper,tightenlines,balancelastpage,superscriptaddress]{revtex4}
\usepackage{amsmath,amsfonts,amssymb,mathrsfs,bm}
\usepackage{verbatim,psfrag,times,CJK}
\usepackage{graphicx,graphics,color,epsfig}
\usepackage{flafter}

\usepackage{indentfirst}
\begin{document}

\newcommand{\ket}[1]{| #1 \rangle} 
\newcommand{\bra}[1]{\langle #1 |}

\title{Nonlinear phase shifts of light trapped in a two-component Bose-Einstein condensate} 
\author{Collin M. \surname{Trail}}
\email{ctrail@gmail.com}
\affiliation{Institute for Quantum Science and Technology,
	University of Calgary, Calgary, Alberta T2N 1N4, Canada}
\author{Khulud \surname{Almutairi}}
\affiliation{Institute for Quantum Science and Technology,
	University of Calgary, Calgary, Alberta T2N 1N4, Canada}
\affiliation{Department of Physics, King Saud University, Riyadh 11451, Saudi Arabia}
\author{David L. \surname{Feder}}
\affiliation{Institute for Quantum Science and Technology,
	University of Calgary, Calgary, Alberta T2N 1N4, Canada}
\author{Barry C. \surname{Sanders}}
\email{sandersb@ucalgary.ca}
\affiliation{Institute for Quantum Science and Technology,
	University of Calgary, Calgary, Alberta T2N 1N4, Canada}
\affiliation{Hefei National Laboratory for Physical Sciences at Microscale,
	University of Science and Technology of China, Anhui 230026, China}
\date{\today}

\begin{abstract}
We investigate a method for generating nonlinear phase shifts on superpositions of photon number states. The light is stored in a Bose-Einstein condensate via electromagnetically-induced transparency memory techniques. The atomic collisions are exploited to generate a nonlinear phase shift of the stored state. The stored light is then revived with the nonlinear phase shift imprinted upon it. We show that this method can be used as a nonlinear-sign gate in the regime where the Thomas-Fermi and mean-field approximations are valid. We test these approximations using realistic parameters and find that these approximations pass the standard tests for validity in a single-component condensate. However, for the two-component condensates considered here, we find that these conditions are insufficiently strict. We find a stronger set of conditions and show for the same set of parameters that the approximations are invalid.
\end{abstract}

\maketitle

\section{Introduction}

Strong optical nonlinearities are desirable for a variety of applications, but challenging to produce. One solution to this challenge is to couple light to a second strongly interacting system. The strong nonlinear interactions in Bose-Einstein condensates (BEC) have been used previously to generate macroscopic superpositions \cite{Savage99}. Furthermore, light can be stored in a BEC and later retrieved using memory techniques based on electromagnetically-induced transparency (EIT) \cite{Harris97, Hau99, Hau01, Dutton02, Hau04, Hau07, Hau09}. There has been recent interest \cite{Simon11, Rempe12} in generating strong optical nonlinearities by storing light in a BEC and exploiting the strong nonlinear atomic interactions. Rispe, He and Simon (RHS) \cite{Simon11} proposed constructing a quantum gate by storing two single-photon pulses in a BEC and using inter-atomic collisions to generate a large conditional phase shift. We extend this proposal here by considering how the same scheme can be applied to multi-photon states such as coherent states.

It has been shown that it is possible to create a universal quantum computer using only linear optics together with single photon sources and photo-detectors using the Knill-Laflamme-Milburn (KLM) scheme for quantum computation \cite{Milburn00}. In the KLM scheme a two-qubit CNOT gate is generated nondeterministically using a pair of  nonlinear-sign gates (NS). The success probability of the NS gates has an upper bound of $1/4$, resulting in a  $1/16$ probability of successfully generating the CNOT gate. Success is heralded, so the CNOT gate can be performed offline to generate an entangled state, which is then used as a resource for teleporting Clifford gates. This is useful for teleporting the CNOT gate, which would otherwise require nonlinear optics. We show that, when the Thomas-Fermi and mean-field approximations are valid, our scheme can be used to deterministically generate the NS gate.

Here we propose a method for generating nonlinear optical phase shifts by storing a light pulse in a cold cloud of atoms, letting the stored pulse evolve under the nonlinearity generated by the atomic interactions, and reviving the light pulse from the cloud. Using the Thomas-Fermi and mean-field approximations, we find a simple expression for the nonlinear optical phase shift. Using this expression, we show that this method can be used to generate the nonlinear-sign gate. Using a realistic set of parameters, we show that these approximations satisfy the standard tests for validity in a single-component condensate. However, we find that the true conditions for the validity of these approximations are stricter due to the presence of the small stored component, and for the same parameters the approximations fail these more rigorous tests.

\section{Methods}

In this section we describe our system and the requirements on its parameters. We consider an $n$-photon pulse propagating through a Bose-Einstein condensate, which is stored as a collective atomic excitation in the condensate. By neglecting the kinetic energy and taking the mean-field approximation of the larger condensate component we are able to derive the equation of motion for the smaller component of the condensate in which the light is stored. Under these approximations, we find that the stored component evolves under a trapping potential plus self-interaction which are both rescaled to account for the interactions with the larger condensate component it is stored within.

We consider a Bose-Einstein condensate with the level configuration illustrated in Fig.\ 1. Initially all atoms are prepared in the $ \left\vert 1\right\rangle$ state. We couple the $\left\vert 2 \right\rangle $ and $\left\vert 3 \right\rangle $ states using a strong continuous beam of light at the resonance frequency $\omega_{23}$ which illuminates the entire condensate at a fixed intensity. We prepare an $n$-photon single-mode pulse whose carrier frequency $\omega_{13}$ is resonant with the $\left\vert 1 \right\rangle$ to $\left\vert 3 \right\rangle$ transition. The shape of this pulse must be carefully matched to the condensate. As we show, the combined effect of the trap and collisions with the host condensate can be treated as a single ``effective" trap. The pulse is matched to this effective trapping potential so that the collisions between stored atoms can be treated perturbatively. 

\begin{figure}
\includegraphics[width=0.90\columnwidth]{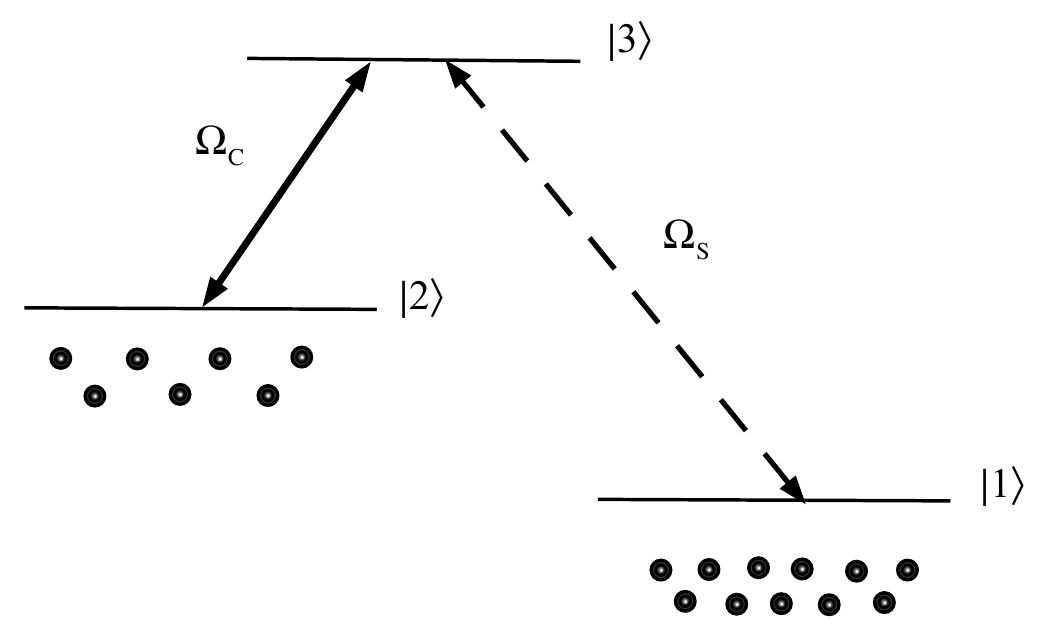}
\caption{
Level scheme for photon storage. The BEC is prepared in level $\left\vert 1\right\rangle$. The signal beam $s$ with Rabi frequency $\Omega_{\mathrm{s}}$ consists of $n$ photons, which are stored as collective excitations in level $\left\vert 2 \right\rangle$.
	}
\label{fig:level}
\end{figure}

We choose the polarization of our coupling beam and of our pulse to minimize coupling to atomic energy levels outside of our three-level system. Here we use the $3S_{1/2}$ levels of sodium-23 used in light storage experiments, which can be modeled as a three-level atomic system \cite{Hau01}. We use the $F=1, m_F=-1$ hyperfine level for $ \left\vert 1\right\rangle$, the $F=2, m_F=1$ hyperfine level for $ \left\vert 2 \right\rangle$, and the $F=2, m_F=0$ hyperfine level for $ \left\vert 3 \right\rangle$. Another possible choice would be to use rubidium-87, for which the loss rates due to atomic collisions would be lower \cite{Dutton02}.

Provided the $n$-photon pulse satisfies the adiabaticity conditions \cite{Dutton02} the atoms stay in the ``dark state" \cite{Arimondo96}. The dark state is the one which does not couple to $\left\vert 3 \right\rangle$, due to destructive interference between absorption of a photon from the coupling beam and absorption of a photon from the $n$-photon pulse. As the pulse enters a region of the condensate, atoms are be transferred from the $\left\vert 1 \right\rangle$ to $\left\vert 2 \right\rangle$ state so that the condensate stays in the local dark state. As all of the atoms of a BEC share the same state, this is a collective excitation of all atoms equally. The dark state does not couple to $\left\vert 3 \right\rangle$, so we can neglect loss due to spontaneous emission as the pulse propagates through the atom cloud. The photons and the excitations of the atoms propagate together and can be treated as quasi-particles called dark-state polaritons. 

If we shut off the coupling beam when the polariton is at the center of the trap, $n$ atoms are collectively transferred from the $\left\vert 1 \right\rangle$ state to the $\left\vert 2 \right\rangle$ state \cite{Lukin00}. This transfer stores the light pulse as a collective excitation in the atomic cloud. We refer to the larger component of the condensate in state $\left\vert 1 \right\rangle$ as the ``host" component, and the smaller component in state $ \left\vert 2 \right\rangle$ as the ``stored" component. 

Both components of the condensate are stored in a trap, 
\begin{equation}
V\left(\bm{x}\right)=\frac{m \omega^2 x^2}{2},
\end{equation}
where $x$ is the distance from the center of the trap to the position $\bm{x}$. For simplicity we have assumed that both components feel an identical trapping potential, which has a simple harmonic form.

The atoms are also subject to collisions within and between components. We initially prepare our condensate in its ground state, so the collisions are low-energy and the s-wave scattering approximation is valid \cite{Sakurai94}. The interaction potential $U_{ij} \left(\bm{x}_1, \bm{x}_2\right)$ between atoms in hyperfine levels $i$ and $j$ simplifies  to $ U_{ij} \delta \left(\bm{x}_1-\bm{x}_2\right)$, where
\begin{equation}
U_{ij}=\frac{4 \pi \hbar^2 a_{ij}} {m}
\end{equation}
and $a_{ij}$ is the s-wave scattering length. The scattering lengths also possess an imaginary component which accounts for loss \cite{Dalgarno97}. Such collisional losses lead to a finite lifetime for the stored component. We restrict ourselves to timescales much shorter than the timescale over which such loss is significant, and can therefore safely neglect loss and ignore the imaginary component of the scattering lengths. We discuss the finite lifetime and possible ways to extend it in Section V.

After the light pulse is stored as a second component of the condensate, the evolution of the stored component is governed by Hamiltonian
\begin{align}
\hat{H}=& \int{{\text{d}^3}} {\bm{x}} \hat{\Psi}_{i}^\dagger \left(\bm{x}, t \right) \Big[ -\frac{\hbar^2}{2m}\nabla^2+V\left(\bm{x} \right)\Big] \hat{\Psi}_{i}\left(\bm{x}, t \right)\nonumber\\
&+\sum_{ij}\int{{\text{d}^3}}{\bm{x}_1} \int{{\text{d}^3}}{\bm{x}_2} \frac{U_{ij} \left(\bm{x}_1, \bm{x}_2\right)}{2} \nonumber  \\
& \times \hat{\Psi}_{i}^\dagger \left(\bm{x}_1, t \right) \hat{\Psi}_{j}^\dagger \left(\bm{x}_2, t \right) \hat{\Psi}_{j} \left(\bm{x}_2, t \right) \hat{\Psi}_{i} \left(\bm{x}_1, t \right),
\end{align}
where $\hat{\Psi}_{i} (\bm{x})$ are the atomic field operators which annihilate an atom in level $i$ at position $(\bm{x})$.

We require that the scattering lengths obey the condition $a_{11} a_{22} > a_{12}^2$, so that we are in the phase-mixing regime rather than the phase-separating regime \cite{Timmermans98}. This is necessary for long storage times, so that the stored component sits at the center of the trap rather than being pushed out to the edge. For the atomic level structure we consider here, the real components of the scattering lengths are $a_{11}=2.75$ nm, $a_{22}=2.85$ nm, and $a_{12}=2.65$ nm \cite{Dutton02}. These scattering lengths satisfy the phase-mixing condition.

Following the approach of RHS \cite{Simon11}, we solve for the ground state of the host condensate component in the mean-field limit and neglect the kinetic energy. For a large number $\left(N \gg 1\right)$ of sufficiently dilute atoms $\left(\frac{N}{V} a^3  \ll 1 \right)$ we can replace the field operators by their expectation values \cite{Stringari99},
\begin{equation}
\psi_i=\left\langle \hat{\Psi}_i \right\rangle.
\end{equation}
We also take the Thomas-Fermi approximation by dropping the kinetic energy term. This approximation is valid when the atomic interactions are sufficiently strong that the cloud radius $R$ is much larger than the characteristic length scale of the trap \cite{Stringari99}, 
\begin{equation}
R \gg \sqrt{\frac{\hbar}{m \omega}}.
\end{equation}
In Section V we show that we introduce error on the order of one part in a thousand into our calculation of the ground state density of the host condensate by neglecting the kinetic energy of the host component and making the mean-field approximation on the host component field operators. We revisit the validity of these approximations for describing the evolution of the smaller stored component in Section V. 

The scattering lengths $a_{11}$ and $a_{12}$ are not equal, so when we store the light pulse we excite some of the host condensate out of its ground state. Because $a_{11}$ and $a_{12}$ differ by only a few percent, and because the stored components we consider here are four to six orders of magnitude smaller than the host component, the effect of the dynamics of the host component on the stored component can be safely neglected. We model the host condensate as though it stays in its ground state after we store the light. 

Under these approximations we find the ground state of the host condensate component, 
\begin{equation}
\left\vert \psi_1 \left(\bm{x}\right) \right\vert^2=\frac{1}{U_{11}}\left(\mu_1-V\left(\bm{x}\right)-U_{12} \left\vert \psi_2 \left(\bm{x}\right) \right\vert^2 \right).
\end{equation}
We substitute Eq.\ (6) into the mean-field equations of motion for the stored component. The dependence of the chemical potential on the number of excitations is small relative to the energy scales considered here, so we can treat it as a constant. We are ultimately interested in the relative phase shift of different components of a superposition of Fock states, which a constant term does not affect, so we can drop the chemical potential term to find that
\begin{align}
i \hbar \frac{\partial \psi_{2}  \left(\bm{x},t\right)}{\partial t} =& \Big[ -\frac{\hbar^2} {2m}\nabla^2+\tilde{V} \left(\bm{x}\right)\nonumber\\
&+ \tilde{U}_{22} \left \vert \psi_{2} \left(\bm{x},t\right) \right\vert^2 \Big] \psi_2 \left(\bm{x},t\right).
\end{align}
The stored component is held by the harmonic trapping potential, but also repelled from the center of the trap by collisions with the host condensate. We model the combined effect of the attraction and repulsion as a single ``effective"  trapping potential
\begin{equation}
\tilde{V}\left(\bm{x}\right)=V\left(\bm{x}\right) \left(1-\frac{U_{12}}{U_{11}} \right)=\frac{m \tilde{\omega}^2 x^2}{2}.
\end{equation}
We chose our states such that  $U_{11}>U_{12}$, so that the stored component remains trapped under the combined effect of the the trap and collisions with the host condensate.

We also see that the effective strength of the collisions between the stored atoms is reduced due to the host atoms they displace. The effective interaction strength, which has been modified to account for the interactions of the stored component with the host component, is
\begin{equation}
\tilde{U}_{22}=\frac{4 \pi \hbar^2 \tilde{a}_{22}} {m},
\end{equation}
where
\begin{equation}
\tilde{a}_{22}=a_{22} -a_{12}^2/a_{11}.
\end{equation}
For the choice of sodium atom levels given above, $\tilde{a}_{22}=0.296$ nm and $\tilde{V}\left(\bm{x}\right)= 0.0364 V\left(\bm{x}\right)$.

We match the shape of the light pulse to the ground state $\phi \left(\bm{x}\right)$ of the effective trapping potential $\tilde{V}$,
\begin{equation}
\phi \left(\bm{x}\right) = \pi^{-3/4} s^{-3/2} \exp\left( -1/2 \left(x/s\right)^2 \right),
\end{equation}
where 
\begin{equation}
s= \sqrt{\frac{\hbar}{m \tilde{\omega}}}
\end{equation}
is the characteristic length scale of the ground-state wave function. This pulse shape allows us to treat the collisions between atoms in the stored component perturbatively, as we show in the next section.
 
In practice the pulse is compressed as it enters the condensate, because the light propagates more slowly as a polariton. Furthermore, the pulse is distorted because the light passing through the lower density region at the edge of the cloud is delayed less than the light passing through the center. This compression and distortion can be compensated for in the preparation of the pulse. 

In deriving the equations of motion we have assumed that the mean-field approximation is valid, but this is certainly not the case for the smaller stored component of the condensate. However, we might expect an analogous set of equations to hold even when the stored component is too small for the mean-field approximation to be valid. If we re-quantize the stored component field operators, we find the equation of motion for the stored component,
\begin{align}
i \hbar \frac{\partial \hat{\Psi}_{2}  \left(\bm{x},t\right)}{\partial t} =& \Big[ -\frac{\hbar^2} {2m}\nabla^2+\tilde{V} \left(\bm{x}\right)\nonumber\\
&+ \tilde{U}_{22} \hat{\Psi}_{2}^\dagger  \left(\bm{x}\right) \hat{\Psi}_{2} \left(\bm{x}\right) \Big] \hat{\Psi}_2  \left(\bm{x},t \right).
\end{align}
We see that the stored component evolves under the combined influence of a harmonic trap and the collisions between the stored atoms, both of which have been rescaled to include the effect of the host component of the condensate.

\section{Two-Photon Case}

When multiple photons are stored in the condensate, collisions between the atoms of the stored component lead to a phase shift. The magnitude of this phase shift depends upon the number of stored photons. 

RHS \cite{Simon11} considered the cross-phase modulation of two photons in different modes, which were stored in collective excitations of a condensate using two different levels of a five level system. In contrast to their work we consider here the self-phase modulation of two photons in a common mode, stored as two collective excitations in the same level of a three level system. 

In this section we calculate the relative shift between the case where two photons are stored, and the zero and one atom cases where such collisions are absent. We derive the relative phase shift using the equation of motion found in the previous section and compare it with the results of RHS \cite{Simon11}.

We first consider the evolution of a two-particle stored component defined as
\begin{equation}
	\psi \left(\vec{\bm{x}},t\right)
		:= \left\langle 0 \right\vert \hat{\Psi}_{2} 
		    \left(\bm{x}_{1}, t\right) \hat{\Psi}_{2} 
		    \left(\bm{x}_{2}, t\right) \left\vert 
		     \Phi_2 \right\rangle,
\end{equation}
where $\left\vert 0 \right\rangle$ is the vacuum state, $\vec{\bm{x}}$ is the vector containing the two coordinate triplets $\bm{x}_{1}$ and $\bm{x}_{2}$,  and
\begin{align}
\left \vert \Phi_2 \right \rangle =& \int {{\text{d}^3}} {\bm{x}_1'} {\text{d}^3}{\bm{x}_2'} \frac{1}{\sqrt{2}} \phi \left(\bm{x}_1'\right) \phi \left(\bm{x}_2'\right) \nonumber \\
&\times \hat{\Psi}_{2}^\dagger \left(\bm{x}_1',0\right)  \hat{\Psi}_{2}^\dagger \left(\bm{x}_2',0\right) \left\vert 0 \right\rangle
\end{align}
is the state with two collective excitations in level  $\left\vert 2 \right\rangle$. 
The evolution of the joint state is
\begin{align}i \hbar \frac{\partial}{\partial t} \psi \left(\vec{\bm{x}}, t\right)  =&\bigg[-\frac{\hbar^2}{2m}\nabla^2_1-\frac{\hbar^2}{2m}\nabla^2_2 +\tilde{V}\left(\bm{x_1}\right) \nonumber \\
&+\tilde{V}\left(\bm{x_2}\right)+\tilde{U}_{22}\delta \left(\bm{x}_{1}- \bm{x}_{2}\right) \bigg] \nonumber \\
& \times \psi \left(\vec{\bm{x}}, t \right).
\end{align} 

We have chosen the shape of the light pulse so that the corresponding atomic excitation matches the ground state of the effective harmonic trap. In the absence of the collision term this would be the ground state of the stored component. We find the evolution of the stored component by treating the collision term as a perturbation to a collision-less system. Since the pairwise interactions between atoms in the state $ \vert 2\rangle$ are much weaker than the effective trapping potential, to good approximation the trapped component remains in the ground state. The effect of the atomic collisions is to shift the energy of the stored component. We can find this energy shift using first-order perturbation theory \cite{Sakurai94}. 

We find the nonlinear phase shift  $\triangle \phi$ due to collisions over interaction time $t$,
\begin{equation}
\triangle \phi := \frac{\triangle E t}{\hbar},
\end{equation}
where
 \begin{align}
 \triangle E=&\left \langle \Phi_2 \right \vert \int{{\text{d}^3}}{\bm{x}} \frac{\tilde{U}_{22}}{2} \hat{\Psi}_{2}^\dagger \left(\bm{x}\right) \hat{\Psi}_{2}^\dagger \left(\bm{x}\right) \hat{\Psi}_{2} \left(\bm{x}\right) \hat{\Psi}_{2} \left(\bm{x}\right)  \left\vert \Phi_2 \right\rangle \nonumber \\
 =& \left\langle 0 \right\vert \int{{\text{d}^3}}{\bm{x}'_1} \int{{\text{d}^3}}{\bm{x}'_2} \int{{\text{d}^3}}{\bm{x}} \int{{\text{d}^3}}{\bm{x}_1} \int{{\text{d}^3}}{\bm{x}_2} \nonumber \\
&\times \hat{\Psi}_{2} \left(\bm{x}'_2\right) \hat{\Psi}_{2} \left(\bm{x}'_1\right)  \hat{\Psi}_{2}^\dagger \left(\bm{x}\right) \hat{\Psi}_{2}^\dagger \left(\bm{x}\right) \nonumber \\
 &\times \hat{\Psi}_{2} \left(\bm{x}\right) \hat{\Psi}_{2} \left(\bm{x}\right) \hat{\Psi}_{2}^\dagger \left(\bm{x}_1\right) \hat{\Psi}_{2}^\dagger \nonumber \\
 &\times \left(\bm{x}_2\right) \frac{\tilde{U}_{22}}{4} \phi  \left(\bm{x}'_1\right) \phi  \left(\bm{x}'_2\right) \phi  \left(\bm{x}_1\right) \phi  \left(\bm{x}_2\right) \left\vert 0 \right\rangle \nonumber \\
 =& \int{{\text{d}^3}} {\bm{x}}  \phi^4 \left(\bm{x}\right)   \tilde{U}_{22}=\left(2 \pi \right)^{-3/2} \tilde{U}_{22} s^{-3}
 \end{align}
 is the perturbation to the energy of the stored component due to the inter-atomic collisions.
 
We can see from this expression that the phase shift on the stored component grows with a stronger effective interaction between the stored atoms and tighter confinement of the stored component. The phase shift has the same form as that found by RHS \cite{Simon11}, with the difference that the effective interaction strength $\tilde{U}_{22}$ depends upon the strength of the interactions between two atoms in the same level rather than the collisions between two atoms in different levels. 

\section{Multi-Photon Case}
In this section we first find the phase shift for an $n$-photon Fock state input to show that it grows nonlinearly with the photon number. Defining the $n$-particle wave function as 
\begin{equation}
	\psi \left(\vec{\bm{x}},t\right)
		:= \left\langle 0 \right\vert \hat{\Psi}_{2} 
		    \left(\bm{x}_{1}, t \right) ,\ldots \hat{\Psi}_{2} 
		    \left(\bm{x}_{n}, t \right) \left\vert 
		     \Phi_n \right \rangle ,
\end{equation}
where $\vec{\bm{x}}$ is now a vector containing $n$ coordinate triplets, we find that it evolves according to the equation
\begin{align} i \hbar \frac{\partial}{\partial t} \psi \left(\vec{\bm{x}}, t \right)=&\bigg[ \sum_{i} -\frac{\hbar^2} {2m}\nabla^2_i+\sum_{i} \tilde{V}\left(\bm{x}_i\right)\nonumber\\
&+\sum_{i<j}  \tilde{U}_{22} \delta \left(\bm{x}_{i}-\bm{x}_{j}\right) \bigg] \nonumber \\
& \times \psi \left(\vec{\bm{x}}, t\right).
\end{align}
As in the two-photon case, we have chosen our atomic state to be the ground state of the effective trapping potential, so the interaction term leads to an additional phase shift. The total energy shift for the $n$-photon case is
 \begin{align}
 \triangle E_n=& \sum_{i<j}^{n}  \int{{\text{d}^3}} {\bm{x}_i} \int{{\text{d}^3}} {\bm{x}_j}  \phi^2 \left(\bm{x}_i\right)  \phi^2 \left(\bm{x}_j\right)\nonumber \\
 & \times \tilde{U}_{22} \delta \left(\bm{x}_i- \bm{x}_j\right) \nonumber \\
  =& \left(2 \pi \right)^{-3/2} \frac{n^2-n}{2} \tilde{U}_{22} s^{-3}. 
\end{align}
We define the nonlinear interaction strength 
\begin{equation}
\hbar \Omega :=\frac{\tilde{U}_{22}}{2 \left(2 \pi \right)^{3/2} s^{3}}   
\end{equation}
such that  
\begin{equation}
\triangle E_n=\left(n^2-n\right) \hbar \Omega.
\end{equation}
We find that the phase shift contains both a nonlinear and a linear term, and that, as in the $2$-photon case, it increases with stronger effective interactions between the stored atoms and with tighter confinement of the stored component. 

\section{Superpositions and Nonlinear-Sign Gate}

Thus far we have only considered Fock state inputs. We now apply the preceding results to the more general case where we inject coherent superpositions of Fock states. We then show how a superposition of zero, one, and two photons leads to a NS gate.

Since an $n$-photon pulse is stored as $n$ collective atomic excitations the final state can be found by applying the appropriate phase shift to each Fock basis component of the complete state,
\begin{equation}
\sum_n c_n \left(t \right)  \left\vert \Phi_n \right\rangle= \sum_n \text{e}^{-i \frac{\triangle E_n}{\hbar} t } c_n \left(0\right)  \left\vert \Phi_n \right\rangle.
\end{equation}
We have ignored loss due to inelastic collisions in our treatment, but in practice such collisions limit the lifetime of the coherence between superpositions of different numbers of collective excitations.

We can estimate the lifetime $\tau$ of the stored component due to collisions with the host component by solving
\begin{equation}
\text{e}^{-i \frac{L}{\hbar} \tau }=1/2
\end{equation}
where 
\begin{equation}
L=\mbox{Im}(U_{12}) 4 \pi  \int_0^\infty \text{d}r r^2 \phi^2 \left( r \right) \psi_1^2 \left( r \right) .
\end{equation}
Using the imaginary component of $U_{12}$ given in \cite{Dutton02} we find that the lifetime is roughly a quarter of a millisecond. This value is in good agreement with the experiments which, for the same choice of atom and level structure considered here, found collisions limit the lifetime to about one millisecond \cite{Hau01}.

It has been found that for a different choice of atomic levels excitations, a bias magnetic field can be tuned to minimize the loss rates, and a light pulse can be stored for up to one second \cite{Hau09}. As discussed previously, gains in lifetime may also be possible by using another atom such as rubidium-87, where the loss rates are typically two orders of magnitude lower \cite{Dutton02}.

We consider the special case of a superposition of up to two stored photons,
\begin{equation}
c_0 \left(0 \right) \left\vert \Phi_0 \right\rangle + c_1 \left(0 \right) \left\vert \Phi_1 \right\rangle + c_2 \left(0 \right) \left\vert \Phi_2 \right\rangle.
\end{equation}
We choose our interaction time $t$ such that $\Omega t =\pi$. For the parameters considered here the pulse would need to be stored for 50 minutes, much longer than the one second storage which has been experimentally demonstrated \cite{Hau09}. However, as discussed in RHS \cite{Simon11} it may be possible to use Feshbach resonances and variable trapping strengths to maximize the interaction strength, significantly reducing the required time. Then we find
\begin{align}
& c_0 \left(t \right) \left\vert \Phi_0 \right\rangle + c_1 \left(t \right) \left\vert \Phi_1 \right\rangle + c_2 \left(t \right) \left\vert \Phi_2 \right\rangle \nonumber \\
= & c_0 \left(0 \right) \left\vert \Phi_0 \right\rangle + c_1 \left(0 \right) \left\vert \Phi_1 \right\rangle - c_2 \left(0 \right) \left\vert \Phi_2 \right\rangle.
\end{align}
This is the nonlinear-sign gate considered in the KLM scheme \cite{Milburn00}.

After performing a NS gate, we would wish to verify that we were successful. By repeating our procedure many times and performing measurements on the output light, we can reconstructing the nonclassical light state using the technique of optical homodyne tomography \cite{Raymer09}. 

We also note that it should be possible to generate optical cat states through a variation of the approach considered by Yurke and Stoler \cite{Yurke86}, by using a coherent state input and generating the optical nonlinearity through atomic interactions as described above.

\section{Critical Analysis}

The success of our method of generating nonlinear interactions depends upon precisely matching the shape of the light pulse to the effective trap ground state. If this is done correctly the subsequent evolution is entirely due to the non-linear perturbation. We had to make approximations to show that the effect of the host condensate component on the stored condensate component could be accounted for by replacing the trap by a rescaled effective trap. It is therefore appropriate to investigate whether the approximations used in deriving our equations for the evolution of the stored component, though valid in describing the gross behavior of the host condensate, significantly modify the dynamics of the stored component. In this section, we first find the conditions under which Eq.\ (13) is valid. We find that the Thomas-Fermi and mean-field approximations must both be valid for Eq.\ (13) to hold. Next, we find the conditions under which these approximations hold, and calculate whether or not these conditions hold for the parameters considered here. 

We start from the second quantized Hamiltonian for a two-component condensate,
\begin{align}
\hat{H}=& \int{{\text{d}^3}} {\bm{x}} 
    \hat{\psi}_{i}^\dagger \left(\bm{x}\right)    
	   \Big[ -\frac{\hbar^2} 
           {2m}\nabla^2+
            V \left(\bm{x}\right)  \Big] \hat{\Psi}_{i}
              \left(\bm{x}\right)\nonumber\\&
               +\sum_{ij}\frac{U_{ij}}{2} 
                   \int{{\text{d}^3}}
                     {\bm{x}}\hat{\Psi}_{i}^\dagger 
                     \left(\bm{x}\right)\hat{\Psi}_{j}^\dagger 
                     \left(\bm{x}\right)\hat{\Psi}_{j} 
                     \left(\bm{x}\right) \hat{\Psi}_{i} 
                     \left(\bm{x}\right),     	    
\end{align}
from which we obtain the equations of motion,
\begin{align}
	i \hbar \frac{\partial \hat{\Psi}_{1} 
	  }{\partial t}
		=& \Big[ -\frac{\hbar^2}
		   {2m}\nabla^2+V
		  + U_{11}  
		\hat{\Psi}_{1}^\dagger 
		   \hat{\Psi}_{1} \nonumber\\
		&+ U_{12}  
		\hat{\Psi}_{2}^\dagger 
		   \hat{\Psi}_{2} 
		      \Big] 
		       \hat{\Psi}_1,
		       \label{eq:Psi1}
		       \\
i \hbar \frac{\partial \hat{\Psi}_{2} 
  }{\partial t}
	=&\Big[ -\frac{\hbar^2}
	   {2m}\nabla^2+V
	    + U_{12}  
		\hat{\Psi}_{1}^\dagger 
		   \hat{\Psi}_{1} \nonumber\\
		&+ U_{22}  
		\hat{\Psi}_{2}^\dagger 
		   \hat{\Psi}_{2} \Big] 
		       \hat{\Psi}_2.
		        \label{eq:Psi2}
\end{align}

For the ground state of the host condensate we find
\begin{equation}
\mu_1 \hat{\Psi}_{1} = \Big[-\frac{\hbar^2}{2m}\nabla^2 +V+ U_{11} \hat{\Psi}_{1}^\dagger \hat{\Psi}_{1} + U_{12}  \hat{\Psi}_{2}^\dagger \hat{\Psi}_{2} \Big] \hat{\Psi}_1.
\end{equation}
Comparing Eq.\ (13) with the exact expression in Eq.\ (32), we see that Eq.\ (13) holds only when the substitution
\begin{equation}
\hat{\Psi}_{1}^\dagger \hat{\Psi}_{1} = \frac{\mu_1 - V- U_{12} \hat{\Psi}_{2}^\dagger \hat{\Psi}_{2} }{U_{11}}
\end{equation}
is valid. We can see from Eq.\ (33) that it is necessary to drop the kinetic energy term by making the TF approximation for this condition to hold. Furthermore, we must be able to drop a factor of $\hat{\Psi}_{1}$ from both sides of Eq.\ (33), which is only valid in the mean-field limit, when the operator $\hat{\Psi}_{1}$ can be replaced by its mean value. Therefore both the TF and mean-field approximations on the host condensate component must be valid for Eq.\ (13), and the results that follow from it, to hold.

We first consider the validity of using the Thomas-Fermi (TF) approximation on the host condensate component. This approximation is known to fail at the edge of the cloud \cite{Stringari96, Feder98}, but we are concerned with the behavior at the center of the trap where the light pulse is stored. 

We test whether or not the TF approximation is self-consistent by evaluating the TF host condensate density profile in the absence of the second component,
\begin{equation}
\left\vert  \psi_1 \left(\bm{x}\right) \right\vert ^2=\frac{\mu_1-V\left(\bm{x}\right)}{U_{11}},
\end{equation}
and calculating its kinetic energy per atom,
\begin{equation}
K \left(\bm{x}\right)=-\frac{\hbar^2}{2m \psi_1 \left(\bm{x}\right)}\nabla^2 \psi_1 \left(\bm{x}\right).
\end{equation}
We account for the effect of the kinetic energy on the host condensate density by including the energy correction term $K$,
\begin{equation}
\left\vert  \psi_1 \left(\bm{x}\right) \right\vert ^2 + \Delta \left\vert  \psi_1 \left(\bm{x}\right) \right\vert ^2=\frac{\mu_1-V\left(\bm{x}\right)-K \left(\bm{x}\right)}{U_{11}},
\end{equation}
which gives us the density correction term,
\begin{equation}
\Delta \left\vert  \psi_1 \left(\bm{x}\right) \right\vert ^2 =\frac{-K \left(\bm{x}\right)}{U_{11}}.
\end{equation}
With this expression for the correction to the host condensate density, we can calculate the correction to the inter-component interaction term appearing in the equation of motion for the stored component,
\begin{equation}
U_{12} \Delta \left\vert  \psi_1 \left(\bm{x}\right) \right\vert ^2= -K \left(\bm{x}\right) \frac{U_{12}}{U_{11}}.
\end{equation}
We can compare this term, $K$ rescaled by the ratio of interaction strengths $\frac{U_{12}}{U_{11}}$, to the stored component's self-interaction energy per atom,
\begin{equation}
U \left(\bm{x}\right)= U_{22} \left\vert  \psi_2 \left(\bm{x}\right) \right\vert ^2,
\end{equation}
in the region where the components overlap. If the rescaled kinetic energy term is smaller then we can safely neglect it. 

In Fig.\ 2, we plot the kinetic energy per atom of the TF solution for the host condensate ground state against the other contributions to the energy of the host condensate. For the plot we consider $10$ stored condensate atoms and $10^6$ host condensate atoms, using a trapping frequency of $100 \pi$ Hz, the atomic mass of sodium 
\begin{equation}
m=3.82 \times 10^{-26} \mbox{kg},
\end{equation}
and scattering lengths given previously. In trap units, we have a trap energy of 
\begin{equation}
E=\hbar \omega=3.30 \times 10^{-32} \mbox{J}
\end{equation}
and an oscillator length scale of
\begin{equation}
d=\sqrt \frac{\hbar}{m \omega}=2.96 \times 10^{-6} \mbox{m}.
\end{equation}
We find that the kinetic energy of the host condensate is smaller than the self-interaction energy of the host component everywhere except at the edge of the condensate. In the absence of a second component, this would be sufficient for showing that the TF approximation were valid (except at the edge). To test whether the approximation is valid for describing the behavior of the stored component, we compare the rescaled kinetic energy to the nonlinear interaction energy, as shown in Fig.\ 3. We find that even with the weighting by $\frac{U_{12}}{U_{11}}$ the kinetic energy term is much larger than the self-interaction energy of the store component, and therefore that the TF approximation on the host component is invalid for modeling the evolution of the stored component.

 \begin{figure}
\includegraphics[width=0.90\columnwidth]{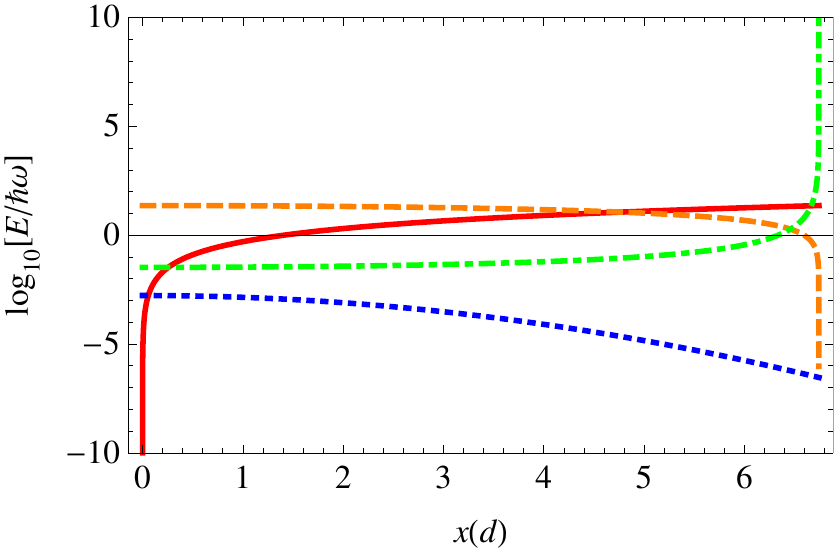}
\caption{
(Color online) Logarithm of energy per host condensate atom in trap energy units vs distance from center of trap in oscillator length units. The solid (red online) line shows the contribution of the trapping potential, the dashed (orange online) line the contribution of collisions with other host condensate atoms, the dotted (blue online) line collisions with stored atoms, and the dash-dotted (green online) line the kinetic energy correction.}
\label{fig:2}
\end{figure}

 \begin{figure}
\includegraphics[width=0.90\columnwidth]{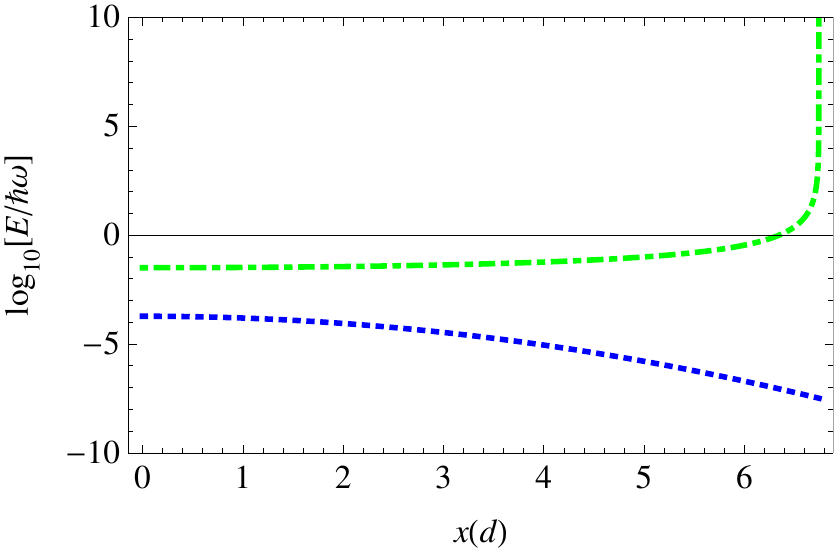}
\caption{
(Color online) Logarithm of energy per atom in trap energy units vs distance from center of trap in oscillator length units. The dotted (blue online) line gives the energy due to collisions of the stored atoms with other stored atoms using the effective interaction strength $\tilde{U}_{22}$, while the dash-dotted (green online) line gives the kinetic energy correction rescaled by $\frac{U_{12}}{U_{11}}$.}
\label{fig:3}
\end{figure}

Next, we consider the validity of the mean-field approximation for the host condensate.  In a mean-field treatment, we replace the atomic field operators by their expectation value. The quantum fluctuation operator is defined as the difference between the exact atomic operator and its mean-field approximation,
\begin{equation}
\delta \hat{\Psi}_{1}: =\hat{\Psi}_{1} - \left \langle \hat{\Psi}_{1} \right \rangle.
\end{equation}
We test the validity of the mean-field approximation by calculating the correction to the mean-field density, known as the quantum depletion, 
\begin{equation}
\left\langle \delta \hat{\Psi}_{1}^\dagger \delta \hat{\Psi}_{1} \right\rangle= \left\langle \hat{\Psi}_{1}^\dagger \hat{\Psi}_{1} \right\rangle - \psi_1^2.
\end{equation}
For the mean-field approximation to be valid, the difference in the dynamics of the stored component cannot depend upon whether or not we include the quantum depletion. For the parameters we consider the interaction strengths are all of the same order, so this is the case if the quantum depletion is smaller than the stored component density. Using the local density approximation \cite{Timmermans97}, we calculated the local magnitude of the depletion. In Fig.\ 4 we compare the density of the depleted host condensate atoms with the density of the condensed atoms for both the host and stored components, using the same parameters as Fig.\ 2. We find the density of the depleted atoms is smaller than the density of the host component, but larger than the density of the stored component. So the mean-field approximation is not valid for describing the behavior of the stored component. 

 \begin{figure}
\includegraphics[width=0.90 \columnwidth]{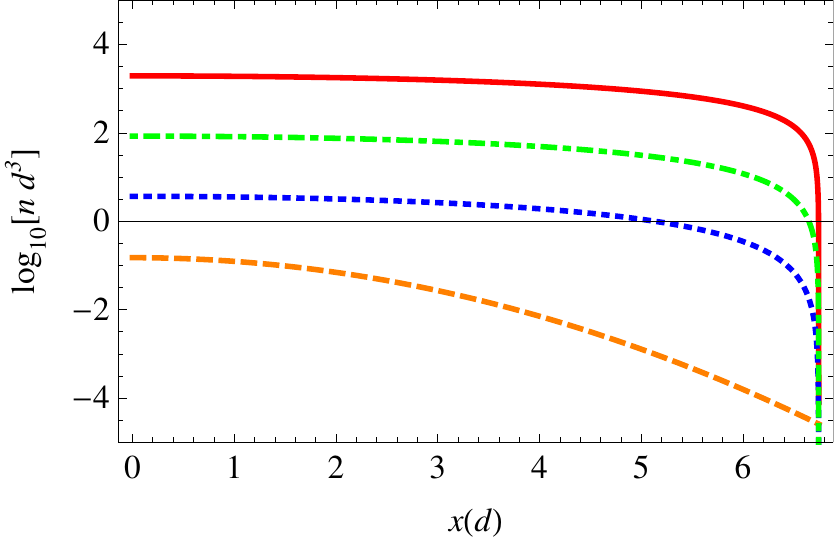}
\caption{
(Color online) Logarithm of atom number per oscillator volume ($d^3$)  vs distance from center of trap in oscillator length units. The solid (red online) line gives the density of the host condensate atoms, the dashed (orange online) line the density of the stored component atoms, the dotted (blue online) line the density of the non-condensed (depleted) host component atoms, and the dash-dotted (green online) line the standard deviation of the density of the host condensate.}
\label{fig:4}
\end{figure}

We can also calculate the variance of the host condensate density to test whether or not it is significant relative to the nonlinear interaction. Expanding the variance
\begin{equation}
\left\langle \left(\hat{\Psi}_1^\dagger \hat{\Psi}_1\right)^2 \right\rangle - \left\langle \left(\hat{\Psi}_1^\dagger \hat{\Psi}_1\right) \right\rangle^2
\end{equation}
in terms of the mean-field and fluctuation operator, we find that the dominant term is proportional to the depletion. In Fig.\ 4 we compare the fluctuations of the host condensate density to the density of the stored component, and find the fluctuations are too large to be ignored when describing the behavior of the stored component.

For both of the approximations we tested, the error introduced was several orders of magnitude larger than the stored component terms. Although the error was small compared to the host component terms, we are concerned here with the much smaller scale of the stored component dynamics. The approximations are incompatible with the required degree of precision for our calculation. Furthermore, the results for the host component density variance suggest that the variability in the phase shift due to collisions between the stored and host component ``wash-out" the nonlinear phase shift. The negative effects of the quantum depletion could be reduced if the coupling between the condensates could be weakened, for example via Feshbach resonances, or if the depletion could be reduced in the region where the condensates overlap. See \cite{Simon14} for an example of the second approach, where the trapping strength of the stored component is made much stronger than the host component, keeping the nonlinear interaction strong while simultaneously making the depletion low.

\section{Conclusions}

We found that an extension of previous results using standard approximations suggests that by storing a coherent state as a collective atomic excitation in a BEC we can generate nonlinear interactions, which allows us to create nonclassical states of light. By using the Thomas-Fermi and mean-field approximations, we were able to find an expression for the ground state of the host component of the condensate. By substituting this expression into the equations of motion for the stored component of the condensate, we found that it evolved under the combined influence of a rescaled trapping potential and a nonlinear self-interaction. By matching our light pulse to the trap ground state, we could treat the self-interaction term as a perturbation, which led to a nonlinear phase shift. We showed that this phase shift could be used to generate the nonlinear-sign gate of the KLM scheme.

We tested the validity of the Thomas-Fermi approximation for our scheme by comparing the kinetic energy of the host condensate ground state solution to the self-interaction energy of the stored component. As we can see in Fig.\ 2, the kinetic contribution to the energy of the host component is much smaller than the combined contribution of the trap and collisions between host atoms, except at the edge of the cloud where the TF approximation is expected to break down. Since the stored component sits at the center of the trap, the edge effects should not be an issue. However, we also see that even at the center of the trap the kinetic contribution is large compared to the interaction between the host and stored component of the condensate. In Fig.\ 3, we have rescaled the host component kinetic energy to account for the fact that it does not directly influence the evolution of the stored component, but rather introduces error into the calculation of the host component profile. This error propagates into the expected evolution of the stored component through the interactions between the two components. Comparing the rescaled kinetic energy to the energy due to collisions between stored component atoms, we see that the kinetic energy term is several orders of magnitude larger even at the center of the trap. Our expression for the evolution of the stored component of the condensate was derived under the assumption that the kinetic energy was negligible, but this result suggests that the error introduced by neglecting the kinetic energy could be many times larger than the self-interaction which gives rise to the nonlinear dynamics.

We tested the validity of the mean-field approximation on the host condensate by comparing local quantum depletion to the stored component density. In Fig.\ 4 we see that the quantum depletion, though much smaller than the mean-field density of the host component, is much larger than density of the stored component. Thus the interactions of the stored component with the depleted fraction of the host component have a greater influence on their dynamics than the interactions within the stored component responsible for the nonlinear dynamics. Again we find that the approximations made in the derivation of our equations of motion for the stored component are too coarse relative to the magnitude of the nonlinear dynamics. As a further test of the validity of the mean-field approximation, we used the quantum depletion to calculate the magnitude of the density fluctuations of the host component. As we see in Fig.\ 4, the fluctuations were also much larger than stored component. The large fluctuations are another indicator that the use of the mean field approximation in our derivation is not valid.

In this paper we have performed a critical analysis of a promising scheme for generating nonlinear optical phase shifts. We found that the validity tests of the TF and MF approximations for a single-component condensate give misleading results when applied to a two-component condensate. Here we have done the necessary work of generalizing these tests for application to a two-component condensate. In doing so, our critical analysis sets a benchmark for other investigations of effecting nonlinear optical transformations in a BEC.

We thank C. Simon and H.\-W. Lau for valuable discussions. This work was supported by CIFAR, NSERC, AITF, and the China Thousand Talent Program.

 \bibliography{Bibfile}{}
\end{document}